\documentclass[a4paper,fleqn]{cas-sc}

\usepackage{amsmath,color}
\usepackage{amssymb}
\usepackage{amsthm}
\usepackage{dsfont}
\usepackage{mathrsfs}
\usepackage[labelformat=simple]{subcaption}
\usepackage{amssymb}
\newtheorem{thm}{Theorem}
\newtheorem{dfn}[thm]{Definition}

\newtheorem{prop}[thm]{Proposition}
\newtheorem{ass}[thm]{Assumption}
\newtheorem{remarks}{Remarks}
\newtheorem{assumptions}{Assumptions}
\def\eff{{\rm eff}}
\usepackage{natbib}

\def\GPP{{\rm GPP}}

\def\tsc#1{\csdef{#1}{\textsc{\lowercase{#1}}\xspace}}
\tsc{WGM}
\tsc{QE}
\tsc{EP}
\tsc{PMS}
\tsc{BEC}
\tsc{DE}

\begin{document}
\let\WriteBookmarks\relax
\def\floatpagepagefraction{1}
\def\textpagefraction{.001}
\shorttitle{Impact of asymptomatic COVID-19 carriers}
\shortauthors{W. Pang et~al.}

\title [mode = title]{Impact of asymptomatic COVID-19 carriers on pandemic policy outcomes}

\author[]{WEIJIE PANG}[
                         orcid=0000-0002-7666-8248]
\cormark[1]

\ead{pangw6@mcmaster.ca}

\address[]{Mathematics \& Statistics, McMaster University, 1280 Main St. West, Hamilton, Ontario, L8S 4L8, Canada}

\author[]{HASSAN CHEHAITLI}[
							orcid=0000-0002-8927-4926]

\author[]{T. R. HURD}[%
                       orcid=0000-0003-1131-5502]

\cortext[cor1]{Corresponding author}

\begin{abstract}
This paper provides a mathematical model to show that the incorrect estimation of $r$, the fraction of asymptomatic COVID-19 carriers in the general population, can account for much of the world's failure to contain the pandemic in its early phases.  The SE(A+O)R model with infectives separated into asymptomatic and ordinary carriers, supplemented by a model of the data generation process, is calibrated to standard datasets for several countries. It is shown that certain fundamental parameters, notably $r$, are unidentifiable with this data. A number of potential types of policy intervention are analyzed. It is found that the lack of parameter identifiability implies that only some, but not all, potential policy interventions can be correctly predicted. In an example representing Italy in March 2020, a hypothetical optimal policy of isolating confirmed cases that aims to reduce the basic reproduction number of the outbreak to $R_0=0.8$ assuming $r=10\%$, only achieves $R_0=1.4$ if it turns out that $r=40\%$.
\end{abstract}

\begin{keywords}
Infectious disease model \sep COVID-19 
\end{keywords}

\maketitle

\section{Introduction}
From the time of its first appearance in late December, 2019 in Wuhan, China, to the time of writing at the end of 2020, COVID-19 continues out of control, and as monitored by \cite{jhdata}  has infected more than 74,000,000 people in more than 185 countries. Delays in implementing strong mitigation policies allowed the disease to spread worldwide. In Italy, while the first two confirmed cases on January 31, 2020 coincided with the government suspension of all flights from China and the announcement of a national emergency, these actions were far from enough to stop the spread of the virus. 

In contrast to other coronaviruses, notably SARS and MERS, an essentially new feature of COVID-19 is the prevalence of asymptomatic infections. This was pointed out  early in the pandemic by \cite{al2020asymptomatic,bai2020presumed,chan2020familial,hu2020clinical,lai2020asymptomatic,tang2020detection,wang2020clinical}. 
\cite{li2020early} mentions that young carriers tend to show milder symptoms than older patients, while generally socializing more. Therefore, including asymptomatic carriers into COVID-19 epidemic models is crucial. Table \ref{rref} shows estimates of $r$, defined to be the proportion of asymptomatic carriers among all viral carriers,  as provided by various studies \cite{wu2020characteristics,mizumoto2020estimating,treibel2020covid,nishiura2020estimation,gudbjartsson2020spread,jia2020modeling,oran2020prevalence}. These estimates of $r$ show no agreement, and consequently our paper aims to study how policy interventions are affected by this uncertainty about the prevalence of asymptomatic carriers. To this end, we modify the standard SIR compartmental ordinary differential equation model introduced in \cite{Kermacketal1927} by splitting the infectious compartment I into  A (asymptomatic carriers) and O (ordinary carriers), and including a compartment E (exposed) to represent the latent phase of the disease before the onset of infectiousness. 
Details about SIR compartmental models  can be found in \cite{bailey1975mathematical,hethcote2000mathematics}. 

Two distinct points of view  are needed to understand infectious diseases such as COVID-19. First we need to understand the disease from the virus point of view. That is, the immunological characteristics of the disease must be determined: How symptoms are manifested, how transmissible the virus is, etc. Equally important is to understand the disease from the perspective of human society, addressing questions of how individual behaviour evolves and adapts to the circumstances of the epidemic. Before a vaccine became available, we had limited means to change the immunological properties of the virus. Instead policy focussed on changing people's behaviour. One of the objectives of this paper is to carefully disentangle the properties of the virus from human behaviour, and to determine how the effect of various policy interventions depends on the presence of asymptomatic carriers. 

Society-wide public interventions may dramatically mitigate the course of the disease if implemented early enough. Unfortunately, despite the example of China where the disease originated, most of the world's countries were not quick or determined enough in their actions. To clarify the effectiveness of possible large scale interventions, we introduce the simplifying assumption that no effective policy measures are undertaken until a certain date, called the policy time $T_{\rm P}$. Our so-called base model is valid prior to $T_{\rm P}$, and represents the course of the disease in its ``natural state'', where society is behaving normally, unconcerned about the COVID dragon that has been stealthily predating on the population.  We imagine that society wakes up suddenly on date $T_{\rm P}$: The focus of this paper is to determine how the effectiveness of various alternative policies depends on the uncertain asymptomatic rate, if they are implemented on that date.

Another point is to distinguish the actual state of the disease at any moment (which is never fully observable) from what we know about the state of the disease at that moment. In other words, in addition to modelling the state of the disease, we need to make assumptions about our observations about the state of the disease. For example, the most publicly available COVID-19 data are daily time series $(C_t, R_t,D_t)$ of the number of active confirmed cases,  removed confirmed cases, and confirmed deaths.  Unfortunately, these numbers are the result of often inefficient data gathering procedures that vary dramatically from country to country. Thus our projections may be very imprecise. In this paper, to enable ``apples to apples'' comparisons, we will assume that each country has in place a system of testing  that generates the time series of $(C_t, R_t,D_t)$. From this daily data, and known studies of COVID and human behaviour, we will infer the {\em actual state} $(S(t), E(t), O(t), A(t), R(t))$ as continuous functions of time $t$ of the disease.

Various kinds of public health policies have been proposed and implemented in different jurisdictions.
\cite{fraser2004factors} recommends the isolation of symptomatic patients and their contacts.   \cite{wu2006reducing} provides recommendations for household-based public health interventions.
\cite{allred2020regional} discusses the effect of asymptomatic patients on the demand for health care. \cite{bousema2014asymptomatic} discusses public health tools that can be used to deal with the presence of asymptomatic carriers. \cite{ferguson2006strategies} apply large scale agent-based simulations to analyze the effect of public health policies on the spread of a virus in its early stages. However, few papers focus on public health policy that targets a viral outbreak whose transmission rate is significantly impacted by asymptomatic  carriers.

The structure of the paper is as follows. Section \ref{sec:model} analyzes the properties of the SE(A+O)R model in which  infectives (I) are separated into asymptomatic carriers (AC) and ordinary carriers (OC). The state process is combined with a model for the observation process in Section \ref{sec:Calib}. The resultant hybrid model is then calibrated to publicly available data for the early stages of the disease in two countries, Canada and Italy, prior to any nation-wide policy implementation.  It is shown that several parameters, notably $r$, are not identifiable in this calibration. In Section \ref{sec:SEIRAP}, we model four types of public health intervention: isolation of infective patients, social distancing, personal protective equipment, and hygiene. It is shown that the outcome of a policy intervention is predictable (not dependent on the unidentifiable parameters) for some of these policies, but not for others. Finally, in Section \ref{sec:diss}, we discuss some principles and pitfalls in designing policy interventions using a model with unidentified parameters.

\section{The Base Model}\label{sec:model}
This paper will extend the standard SEIR (Susceptible, Exposed, Infectious, Removed) ordinary differential equation (ODE) model by splitting the infectious compartment into two disjoint sets, asymptomatic viral carriers (AC) and ordinary carriers (OC). 
The {\em base model} discussed in this section represents the outbreak only in its initial stages, before the first major policy disease-specific intervention such as a country-wide lockdown. Later in the paper, we extend the picture to account for  policy interventions that aim to dramatically alter the progress of the disease. Our model will carefully distinguish between ``actual'' cases and ``confirmed'' cases. Confirmed cases are the result of an observation process applied to, but not impacting, the actual system. 

\subsection{Asymptomatic and Ordinary Carriers} At the time of writing, there is still a large amount of uncertainty about the prevalence of ACs in the general population, partly because of different definitions of what is meant by an AC. We adopt clear cut operational definitions similar to those of \cite{oran2020prevalence} that do not depend on whether or not the case has been tested or otherwise confirmed:
\begin{dfn}
	\label{def:acoc}
	
	\begin{enumerate}
		\item An {\em Asymptomatic Carrier} (AC) is someone who 
		\begin{enumerate}
			\item has been exposed to COVID and  is currently infectious;
			\item  will show no noticeable COVID symptoms for the entire infective period.
			
		\end{enumerate}
		
		\item An {\em Ordinary Carrier} (OC) is someone who 
		\begin{enumerate}
			\item has been exposed to COVID  and  is currently infectious;
			\item  will show some noticeable COVID symptoms  at some point during the entire infective period.
		\end{enumerate}
	\end{enumerate}
	
\end{dfn}

\begin{remarks}
	
Based on the Definition \ref{def:acoc}, presymptomatic carriers and carriers with mild symptoms are included in OC, as long as they eventually show recognizable symptoms. 
\end{remarks}

With Definition \ref{def:acoc},
AC individuals are unlikely to be identified and confirmed, leading to great uncertainty in their prevalance. Moreover, various studies define the term ``asymptomatic'' differently. These intrinsic difficulties make it extremely problematic to determine the key parameter, the {\em asymptomatic fraction} $r$, which we define to be the long-time limiting fraction of individuals who had the disease but were asymptomatic. Note that $r$ is an intrinsic characteristic of the infection mechanism, but its value may differ greatly from studies that adopt a different definition. Table \ref{rref} displays estimates of $r$ made in a number of studies, which we see have sharpened somewhat over 2020, but still remain very uncertain. 

The most conclusive study on Table  \ref{rref}, \cite{oran2020prevalence}, summarizes their important message: ``On the basis of the three cohorts with representative samples--Iceland and Indiana, with data gathered through random selection of participants, and Vo', with data for nearly all residents--the asymptomatic infection rate may be as high as 40\% to 45\%. A conservative estimate would be 30\% or higher to account for the presymptomatic admixture that has thus far not been adequately quantified.''

\begin{table}[width=.7\linewidth,cols=3,pos=h]
\caption{Estimation of asymptomatic rate $r$ of COVID-19 from various researchers between February 2020 to September 2020.} 
\label{rref}
	\centering
	\begin{tabular*}{\tblwidth}{@{} CCC @{} }
		\toprule
		Date & Sources & Estimation of $r$ \\
		\hline\hline
		February 2020 & \cite{novel2020epidemiological} & 1.2\%\\
		\hline
		February 2020 & \cite{mizumoto2020estimating} & 17.9\% (15.5 - 20.2\%) \\
		\hline
		May 2020 & \cite{treibel2020covid} & 1.1\% - 7.1 \% \\
		\hline
		May 2020 & \cite{nishiura2020estimation} & 30.8\% ( 7.7\% - 53.8\%)\\
		\hline
		May 2020 & \cite{arons2020presymptomatic} & 0\% - 6\%\\
		\hline
		June 2020 &\cite{gudbjartsson2020spread} & 5.7\% - 58.3\%\\
		\hline
		August 2020 & \cite{jia2020modeling} & 58.9\% - 92.5 \% \\
		\hline
		September 2020 &\cite{oran2020prevalence} & 40\% - 45\%\\
		\bottomrule	
	\end{tabular*}

\end{table}

\subsection{System Assumptions}\label{sec:asssym}
In the present section, we specify the basic SEAOR model for a fully homogeneous well-mixed population, applicable in a jurisdiction before any significant COVID mitigation response has been initiated. 
\begin{assumptions}
	\begin{enumerate}
		\item The total population $N=S(t)+E(t)+A(t)+O(t)+R(t)$ is constant. The natural birth rate equals the natural (pre-COVID) death rate. No immigration or emigration is considered from other countries.
		
		\item  The removed compartment includes the recovered population who acquire permanent immunity and all COVID deaths. No vaccine is yet available for the virus, which means all people are susceptible prior to their first exposure.  
	\item  The population is homogeneous and well-mixed, and thus the mass-action principle is assumed for the infection transmission.  Both the latent period (exposed and not yet infectious) and infectious period are exponential random times. 
	\end{enumerate}
\end{assumptions}

\begin{remarks} 
It is common modeling practice to extend the ODE approach to allow for $M$ communities, with the homogeneous and well-mixed assumption within each community. It is also common practice to model the latent and infectious periods as random times with a more realistic gamma distribution. 

\end{remarks}

\begin{table}[width=.5\linewidth,cols=2,pos=h]
	\caption{Notation}
	\label{tab:sym}
	\centering
	\begin{tabular*}{\tblwidth}{@{} C|L@{} }
		\toprule
		$S$ & Susceptible population  \\ 
		\hline
		$E$ & Those exposed to COVID-19 but not yet infectious \\ 
		\hline
		$O$ & Ordinary carriers \\ 
		\hline
		$A$ & Asymptomatic carriers \\
		\hline
		$R$ & Removed population\\
		\hline
		$\alpha^O$ & Transmission rate of Category (S) from Category (O)\\
		\hline
		$\alpha^A$ & Transmission rate of Category (S) from Category (A)\\
		\hline
		$\beta$ & Inverse duration time in Category (E)\\
		\hline
		$\gamma^O$& Inverse duration time in Category (O) \\
		\hline
		$\gamma^A$& Inverse duration time in Category (A)\\
		\hline
		$r$& Fraction of (E) that become  (A)\\
		\bottomrule		
	\end{tabular*}

\end{table}

Based on these assumptions and using the notation identified in Table \ref{tab:sym}, the SEAOR model is defined as follows:
\begin{equation}\label{SEAOR}
\begin{split}
\frac{dS}{dt} &= -(\alpha^O  O  + \alpha^A A)S\\
\frac{dE}{dt} &= (\alpha^O  O  +\alpha^A A)S - \beta E\\
\frac{dA}{dt} &= r\beta E - \gamma^A A  \\
\frac{dO}{dt} &= (1-r)\beta E  - \gamma^O O \\
\frac{dR}{dt} &= \gamma^O O +\gamma^A A  \ . \\
\end{split}
\end{equation}

\subsection{Model Parameters} Our objective is to consider targeted interventions that can mitigate the stark outcomes that emerge from the base contagion model. Knowing the meaning of the model parameters is important in understanding the type of data that will be needed to determine them, and how potential policy interventions act. \begin{enumerate}
	\item {\em Transmission parameters} $\alpha$: These parameters are a product of more fundamental parameters that arise in ``microscale''  agent-based models (ABMs) and network models. In general, $\alpha^A,\alpha^O$ are the average daily rate of new exposures that occur, per susceptible, per asymptomatic or ordinary infectious carrier. They are naturally a product of three factors, as described in \cite{hurd2020covid}:
	\begin{itemize}
		\item $\kappa=\kappa^O=\kappa^A$ is a constant describing the average number of significant social relations per individual. Since it is based on studies of normal social conditions, its value does not change under any policy intervention.    
		
		\item $z$ is the daily rate of ``close contacts'' per significant social relation. It naturally changes when an individual becomes symptomatic and so we should expect that $z^A>z^O$. 
		
		\item Infectivity $\tau$ is the probability that a close contact with an infective person actually leads to exposure (hence the disease). This can be reduced by policies that either boost immunity or reduce viral transfer. While one might expect that $\tau^A <\tau^O$, in this paper we will assume the worst case $\tau^A = \tau^O$. 
	\end{itemize}
	\item COVID-19 studies made during the early stages of the pandemic, notably \cite{colizza2007modeling,diekmann2000mathematical}, suggest that the average latent period is about 5 days and average infectious period is 6 days. Under the exponential time assumption, these values justify the estimators for $\beta,\gamma^O$ we will use throughout this paper:
	\begin{equation}\label{eq:betagamma}
	\begin{split}
	\hat\beta &= \frac{1}{\mbox{Latent Period}} = 0.2\ ,\quad
	\hat\gamma^O  = \frac{1}{\mbox{Infectious Period}} =0.167
	\end{split}
	\end{equation} 
\end{enumerate}
In view of the difficulty to observe asymptomatic cases, the parameter $\gamma^A$ will be hard to determine, so it is natural to make the assumption that $\gamma^A=\gamma^O:=\gamma$.  

\subsection{The Reduced Model} 
For the remainder of the paper, we make the assumption that $\gamma^A=\gamma^O:=\gamma$.  With this condition, the SEAOR model can be almost fully understood from properties of the following {\em reduced model}:
\begin{equation}\label{SEIReff}
\begin{split}
\frac{dS}{dt} &= - \alpha^\eff S I\\
\frac{dE}{dt} &= \alpha^\eff S I - \beta E\\
\frac{dI}{dt} &= \beta E  - \gamma I \\
\frac{dR}{dt} &=\gamma I  \\
\end{split}
\end{equation}
where $\alpha^\eff := (1-r)\alpha^O+r\alpha^A$. This fact derives from the following easily proved result. 

\begin{prop}\label{Prop2} Suppose $\gamma^A=\gamma^O:=\gamma$ and the initial conditions for \eqref{SEAOR}  satisfy 
	$A(0)=r(A(0)+ O(0))$. Then the solution of \eqref{SEAOR} is given by the solution of \eqref{SEIReff}  by setting $A(t)=rI(t), O(t)=(1-r)I(t)$ for all $t$. 
\end{prop}

It will turn out that calibration of the reduced model to confirmed daily new case data will determine $\alpha^\eff$, but $\alpha^O,\alpha^A$ will not be determined separately without additional data. If we define $\rho=\alpha^A/\alpha^O=(\tau^Az^A)/(\tau^Oz^O)$, then we have
\begin{equation}
\label{}
\alpha^O=(1-r+r\rho)^{-1}\alpha^\eff,\quad \alpha^A=\rho(1-r+r\rho)^{-1}\alpha^\eff\ .
\end{equation}
In our benchmark models, we will take $\rho=4$, which would result from the plausible relationships $\tau^A=\tau^O, z^A=4z^O$, in other words, if asymptomatic carriers are four times more sociable and similarly infectious compared to symptomatic carriers.

\subsection{Linearized Analysis} The linearization of \eqref{SEAOR} about the disease-free equilibrium  $(N,0,0,0,0)$ provides a useful starting point to understand the early stages of the COVID-19 pandemic. With the restriction $\gamma^A=\gamma^O=\gamma$, this can be reduced to the following 3-d linear system with state vector $X(t)=(E(t),A(t),O(t))'$:
\begin{equation}\label{lin_SEAOR}
\begin{split}
\left(\begin{array}{c} dE/dt \\ dA/dt\\ dO/dt  \end{array}\right)
= \left(\begin{array}{c c c} -\beta & \alpha^A N& \alpha^O N \\ \beta r& -\gamma &0 \\ \beta (1-r)&0& -\gamma \end{array} \right) 
\left(\begin{array}{c} E \\ A\\O\end{array} \right) 
:= B \left(\begin{array}{c} E \\ A\\O\end{array} \right).
\end{split}
\end{equation}
Any solution vector  $X(t)=(E(t),A(t),O(t))'$ of \eqref{lin_SEAOR}  generates an approximate solution of \eqref{SEAOR} by setting \begin{eqnarray}
R(t) & = &R(0)+\gamma \int^t_0 (A(s)+O(s))\ ds\ , \\
S(t) & = & N-E(t)-A(t)-O(t)-R(t) \ .
\end{eqnarray}
The linearized approximation \eqref{lin_SEAOR} will be sufficiently accurate as long as $S(t)/ N$ is sufficiently close to $1$.

The spectral properties of the matrix $B$ can be summarized by the three eigenvalue-eigenvector pairs:
\begin{equation}
\lambda_+\ ,\ V_+=\left(\begin{array}{c} 1 \\ rv_+\\(1-r)v_+\end{array} \right) ;\quad 
\lambda_-\ ,\  V_-=\left(\begin{array}{c} 1 \\ rv_-\\(1-r)v_-\end{array} \right) 
;\quad 
-\gamma \ , \  V_\gamma=\left(\begin{array}{c} 0  \\ 1\\-1\end{array} \right) 
\end{equation}
where \begin{equation}
\label{lambda_v}
\lambda_\pm = -\frac{\beta + \gamma}{2} \pm \sqrt{\frac{(\beta - \gamma)^2}{4} + \alpha^\eff\beta N},\quad 
v_\pm=\frac{\beta+\lambda_\pm}{\alpha^\eff N}\ . 
\end{equation}
Furthermore, the {\em basic reproduction number}  (``R-naught'') is 
\begin{equation}
\label{Rnaught}
R_0=\frac{\alpha^\eff N}{\gamma}\ .
\end{equation}

Our primary interest will focus on cases of a pandemic with $R_0>1$ which is equivalent to  $\lambda_+>0>\lambda_-$. In such situations, the general solution $X(t)=(E(t),A(t),O(t))', t\ge 0$ of \eqref{lin_SEAOR} with any positive initial small COVID infection has the form 
\begin{equation}
\label{Xsolution }
X(t)=a_1e^{\lambda_+ t}V_++a_2e^{\lambda_- t}V_-+a_3e^{-\gamma t}V_\gamma
\end{equation}
for coefficients $a_1,a_2,a_3$, and will exhibit an exponentially fast convergence to a multiple of the dominant eigensolution 
\begin{equation}
\label{Xsolution2}
X_+(t)=(1, rv_+,(1-r)v_+)' e^{\lambda_+ t}\ .
\end{equation} 
We emphasize that this dominant solution $X_+(t)$ corresponds to the early exponentially growing phase of the pandemic. Its  rate $\lambda_+$  depends on the effective transmission rate $\alpha^\eff=(1-r)\alpha^O+r\alpha^A$, rather than on $r,\alpha^O,\alpha^A$ separately. In Section 3 we will find that $X_+(t)$ does indeed provide a reasonably good fit for various countries during the early stages of COVID. 
Note also that by Proposition \ref{Prop2}, $(1, v_+)' e^{\lambda_+ t}$ will be a solution of the linearization of the reduced system \eqref{SEIReff}.

\subsection{Time Periods}
Let us now consider a country, for example Italy or Canada, and let $t=0$ denote time 00:00 on  January 1, 2020. We assume that the ODE model is an acceptable approximation after a time called the {\em pandemic time} $T_1$ when a sufficient number of cases have been generated: We define $T_1$ to be the start of the first day the confirmed cumulative cases exceeded $50$ cases in the country under study.   
Our aim is first to study the short period of time $[T_1,T_2]$ called the {\em pre-policy period}, starting at the pandemic time $T_1$ and ending at the {\em policy time} $T_2$, defined to be the time of the first nation-wide policy intervention. Because policy changes taking place on $T_2$ will take several days to have an observable effect on case numbers, we first calibrate our model to the {\em calibration period} $[T_1,T_2+5]$, which includes 5 days following $T_2$. In Section 4 we will study the effect of possible public health policy interventions implemented at the policy time $T_2$ for the six-weeks long {\em post-policy period} $[T_2,T_3]$ with the {\em end time} $T_3=T_2+42$.

In different countries around the world, the pandemic time $T_1$ and policy time $T_2$ typically occurred in February and March 2020. In this paper, we focus for illustrative purposes on two countries, Italy and Canada, where these dates are summarized in Table \ref{tab:tcountry}. In Canada, the confirmed cumulative cases reached 51 on the day following the pandemic time $T_1 = 48$ (February 18). On the day following the policy time $T_2 = 73$ (March 15), Ontario, Canada's largest province, mandated province-wide public school closures and other provinces made similar announcements. Italy's pandemic time was the same, $T_1 = 52$ (start of February 22), and on that date the confirmed cumulative cases reached 79. On the other hand, Italy imposed a nation-wide lockdown policy on the day following the policy time $T_2 = 68$ (March 10).

\begin{table}[width=.85\linewidth,cols=3,pos=h]
	\caption{The pre-policy and post-policy periods in Italy and Canada extend over $[T_1,T_2]$ and $[T_2,T_3]$ respectively, where $T_1,T_2$ are shown in the table and $T_3=T_2+42$, which is six weeks after policy time. } 
	\label{tab:tcountry}
	\centering
	\begin{tabular*}{\tblwidth}{@{} C|CC@{} }
		\toprule
		Country & Pandemic Time ($T_1$) & Policy Time ($T_2$) \\
		\hline\hline
		Italy &  $T_1=52$   & $T_2=68$ \\
		~ & (start of February 22, 79 Cases) & (start of March 10, Nationwide lockdown)\\
		\hline
		Canada & $T_1=48$   &$T_2=73$\\
		~& (start of February 18, 51 Cases) & (start of March 15, Ontario School Shutdown)\\
		\bottomrule
	\end{tabular*}

\end{table}

\section{Pre-Policy Calibration}\label{sec:Calib} This section will show that the SEAOR model with the parametric restriction  $\gamma^A=\gamma^O:=\gamma$ provides a reasonably good fit to the disease in its early stages for countries such as Canada and Italy, when calibrated to fit the observed data, specifically, the confirmed daily new case data for the period $[T_1,T_2+5]$ (i.e. the pre-policy period plus 5 days). 

\subsection{Measurements and Observations} The ODE system \eqref{SEAOR} captures the dynamics of the unobserved state of the population, and its parameters do not depend on how the system is observed. 
Parameters $\alpha^O ,\alpha^A$ combine characteristics of the disease with the social mixing parameters and behaviour, and hence depend strongly on policy and jurisdiction. 
There are many possible methods for observing the state and evolution of the system. We make some simplifying assumptions about the model and the observation process. 
\begin{assumptions}

During the pre-policy period $[T_0,T_2]$, model parameters are constant. Among the OC population, an expected fraction $\phi^O$ are counted as confirmed cases, typically as a result of either a positive RT-PCR test (``swab test'') or a diagnosis by symptoms. 
In contrast, none of the AC population are counted as confirmed cases ($\phi^A=0$).
\end{assumptions}

\def\DNC{{\rm DNC}}
\def\MSE{{\rm MSE}}
Let us denote the confirmed daily new cases on the day ending at time $T_1+k$ by $\widehat \DNC_k$, for $k=1,2,\dots, K=T_2+5-T_1$. We make the assumption that the data generating process is a random process fluctuating around $X(t)$, the dominant solution \eqref{Xsolution2} of the linearized SEAOR model:
\begin{equation}
\label{DGP}
\widehat \DNC_k =\left( \phi^O(1-r)\beta \int_{T_1+k-1}^{T_1+k} E(s) ds\right)e^{\zeta_k}\ .
\end{equation}
Here $(\zeta_k)_{k=1,2,\dots, K}$ is an i.i.d. $N(0,\sigma^2)$ sequence of residuals, $E(s)=E_0e^{\lambda_+(s-T_1)}$ with $E_0=E(T_1)$, and $\lambda_+$ is given by \eqref{lambda_v}. 

\subsection{Pre-Policy Calibration: Italy and Canada} 

The logarithm of the data generation process \eqref{DGP} is a simple linear regression
\[ \log \widehat \DNC_k =\kappa + \lambda_+ k +\zeta_k, \]
with $\kappa:=\log(\phi^O(1-r)\beta E(T_1)(1-e^{-\lambda_+})/\lambda_+)$. This leads to the least square estimates $\hat\theta=(\hat\kappa,\hat\lambda_+)$  for two identifiable parameters
\begin{equation}
\label{ }
\hat\kappa = \text{intercept coefficient}\ ; \hat\lambda_+= \text{ slope coefficient } \ .
\end{equation} 
Given $\hat\kappa,\hat\lambda_+$,  the factors of  $ e^\kappa=\phi^O(1-r)\beta E_0(1-e^{-\lambda_+})/\lambda_+$ are not separately identifiable. The mean squared regression error $\hat\sigma^2=\MSE_K$ is
\begin{equation}
\label{ }
\hat\sigma^2=\MSE_K= \frac1{K} \sum_{k=1}^K |\log(\widehat \DNC_k)-\kappa-\lambda_+k|^2
\end{equation}

\begin{figure}[h!]
	\begin{subfigure}{.5\textwidth}
		\centering
		\includegraphics[width=0.95\linewidth]{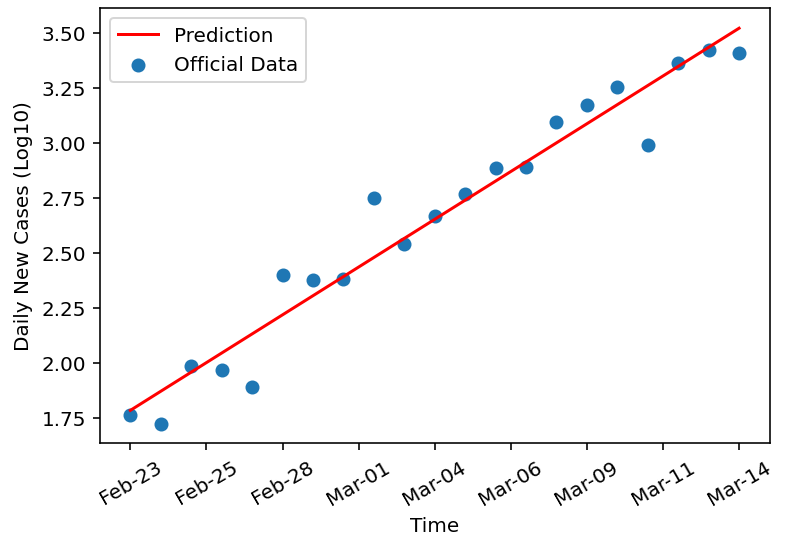}
		\caption{Italy}
	\end{subfigure}%
	\begin{subfigure}{.5\textwidth}
		\centering
		\includegraphics[width=0.95\linewidth]{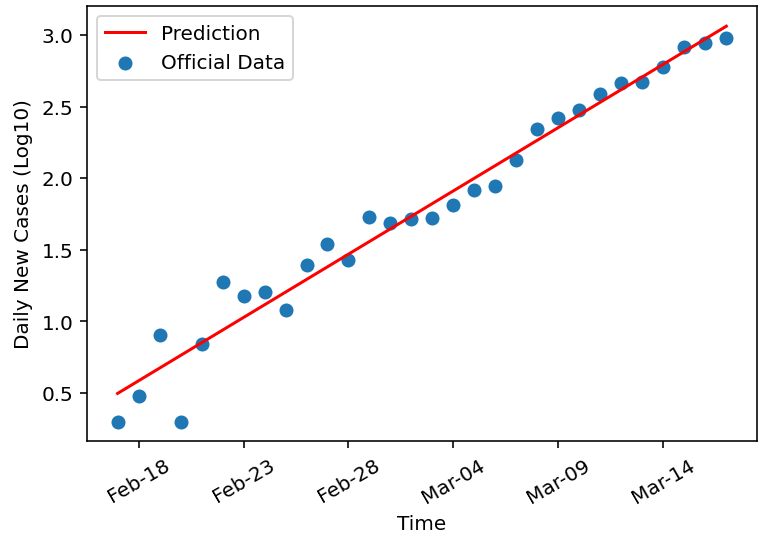}
		\caption{Canada}
	\end{subfigure}
	\caption{Simulation and official data of daily new cases (DNC) for Italy and Canada in the pre-policy calibration period $[T_1,T_2+5]$.}
	\label{fig:alphaeff}
\end{figure}

Figure \ref{fig:alphaeff} displays values given by the calibrated model and observed data for daily new cases on a log-scale. We note that for both Italy and Canada, the calibrated model gives a reasonably good fit, albeit with a significant degree of noise.  Estimators $\hat\beta,\hat\gamma,\hat\lambda_+,\hat\kappa$, when combined with \eqref{lambda_v}, lead to estimates for further important parameters $\hat\alpha^\eff,\hat\lambda_-,\hat v_+, R_0 $. Table \ref{tab:alphaeffitaly} shows those parameter values resulting from the calibration for Canada and Italy. 

\begin{table}[width=.9\linewidth,cols=8,pos=h]
	\caption{Parameter estimates for Italy and Canada during the pre-policy period. } 
	\label{tab:alphaeffitaly}
	\centering
	\begin{tabular*}{\tblwidth}{@{} C|CCCCCCC@{} }
		\toprule
		Country&$\hat\kappa$ & $\hat\lambda_+$ & $\rm MSE = \hat\sigma^2$ &  $\hat\lambda_-$&$\hat\alpha^\eff$ &$\hat v_+$  &$R_0$  \\
		\hline\hline
		Italy &  1.6974 &  0.1999 & 0.015 & -0.5666 & 1.214e-08 & 0.546 & 4.40 \\
		\hline
		Canada & 0.4090  & 0.2037 & 0.021 & -0.5703 & 1.988e-08 &  0.540 & 4.48 \\
		\bottomrule	
	\end{tabular*}
\end{table}

These parameter estimates do not fully determine the model and its initial conditions: $r,\phi^O,\alpha^O,\alpha^A, E_0$ are left undetermined, but constrained by two equations: 
\begin{eqnarray}
\label{hatkappa}
\hat\kappa&=&\log(\beta\phi^O(1-r)E_0(1-e^{-\hat\lambda_+})/\hat\lambda_+)\ ,\\
\label{hatalpha} \hat\alpha^\eff &=& (1-r)\alpha^O+r\alpha^A
\end{eqnarray}
Table \ref{prepolicyCOVID} provides the best-fit values for the actual state of the pandemic on the dates $T_1,T_2$ using the linearized solutions
\begin{eqnarray}
\label{ }
E(t)&=&(\beta\phi^O(1-r)(1-e^{-\hat\lambda_+})/\hat\lambda_+)^{-1} e^{\hat\kappa+\hat\lambda_+(t-T_1)}\,\\ I(t)&=&\hat v_+ (\beta\phi^O(1-r)(1-e^{-\hat\lambda_+})/\hat\lambda_+)^{-1} e^{\hat\kappa+\hat\lambda_+(t-T_1)}
\end{eqnarray} 
assuming the model may have different values of $\phi^O(1-r)$. The removed value $R(t)$ can be accurately approximated by $\gamma\int^t_{-\infty}I(s)\ ds=\gamma I(t)/\hat\lambda_+$. Note that to determine the separated compartment populations $A(t)=rI(t), O(t)=(1-r)I(t)$, we also need the value of $r$.  This indistinguishability of models leads to difficulty about the efficacy of different health policy interventions, as we will investigate in the next section. 

\begin{table}[width=.9\linewidth,cols=8,pos=h]
	\caption{The actual state of the pandemic at times $T_1,T_2$ obtained from the linearized model, depending on the additional parameters $\phi^O,r$. Given $r$, the asymptomatic and ordinary case numbers will be $A(t)=rI(t), O(t)=(1-r)I(t)$.  } 
	\label{prepolicyCOVID}
	\centering
	\begin{tabular*}{\tblwidth}{@{} C|CCCCCCC@{} }
		\toprule
		$\phi^O(1-r)$& Country & $E(T_1)$&$I(T_1)$&$R(T_1)$&$E(T_2)$&$I(T_2)$&$R(T_2)$  \\
		\hline\hline
		20\% &Italy & 150  &  82 & 68 & 8208 & 4478 & 3731    \\
		&Canada&   41 & 22 & 18 & 15279 & 8251 & 5609 
		\\
		\hline
		40\% &Italy & 75 & 41 & 34 &  4104 & 2239 & 1865    \\
		&Canada & 20 & 11 & 9 & 7639 & 4125 & 2804 \\
		\hline
		60\% &Italy &  50 & 27 & 22 & 2736 & 1492 & 1243    \\
		&Canada& 13 & 7 & 6 & 5093 & 2750 & 1869  \\
		\hline
		80\% &Italy &  37 & 20 & 17 & 2052 & 1119 &  932  \\
		&Canada& 10  &  5 & 4 & 3819 & 2062 & 1402    \\
		\hline
		100\% &Italy & 30 & 16 & 13 & 1641 & 895 & 746  \\
		&Canada& 8 & 4 & 3 & 3055 & 1650 & 1121 \\
		\bottomrule
	\end{tabular*}
\end{table}

\section{Public Health Policy Interventions}\label{sec:SEIRAP}

Policy makers seek to mitigate the societal damage from the disease by adopting actions or policies that reduce the transmission parameters $\alpha^O,\alpha^A$, thereby decreasing the exposure rate of the susceptible population to viral carriers. Other important parameters, notably $\beta,\gamma,r$, are not easily controllable and will be taken as unchanged by such policies. 

In this section, we consider the example of Italy under a range of hypothetical scenarios where a substantial health policy intervention is made instantaneously at the policy time $T_2$, with a constant level of effort thereafter. Since the parameters $\phi,r$ are not identifiable from the database used in calibration, we explore how the effectiveness of any policy depends on $\phi, r$.  The non-linear system of ODEs \eqref{SEAOR}  is thus solved for pre-policy parameters for the period $[T_1,T_2 ]$ with the initial condition at time $T_1$ given by Table \ref{prepolicyCOVID} for different values of a product $\phi(1-r)$, and then with post-policy parameters for the period $[T_2,T_3]$.
The starting conditions at time $T_1$ have the form $(S(T_1),E(T_1),A(T_1),O(T_1),R(T_1))$ with $A(T_1)=r(A(T_1)+O(T_1))$, and $\gamma^A=\gamma^O=\gamma$ at all times. Therefore Proposition 2 applies for the period $[T_1,T_2]$ and again for the period $[T_2,T_3]$ and so in the following explorations, we may solve the reduced non-linear system \eqref{SEIReff} and use the equations $A(t)=r I(t), O(t)=(1-r)I(t)$ to obtain the desired solution of \eqref{SEAOR}. 

\subsection{Policy Choices}

To quantify the effect of policy $p$, we first define the {\em maximal effect} of the policy to act on the pair of transmission parameters $(\alpha^O,\alpha^A)$ leading to new values $(1-v^O_p)\alpha^O,(1-v^A_p)\alpha^A$, where $(v^O_p,v^A_p)\in [0,1]^2$ is called the {\em maximal effect vector}. Then, if the policy $p$ is adopted with a {\em partial degree of effort}  $e_p\in [0,1]$, the policy effect on the transmission parameters is assumed to lead to new values 
\begin{equation}
\label{policy}
(\alpha^O,\alpha^A) \longrightarrow (\alpha^O_p, \alpha^A_p) = \left((1-e_pv^O_p)\alpha^O,(1-e_pv^A_p)\alpha^A\right)\ . 
\end{equation}

At a more fundamental level, different policies typically act by directly reducing some of the infectious contact parameters $z^O,z^A$ or the infection probabilities $\tau^O,\tau^A$.  By their definition, $\kappa=\kappa^O=\kappa^A$ are not changed in short term policies.  Under the assumptions of Proposition \ref{Prop2}, the net effect of each policy on the impact of the disease is only through the effective parameter $\alpha^{\eff}=(1-r)\alpha^O+r\alpha^A$ where $\alpha^O=\kappa z^O\tau^O, \alpha^A=\kappa z^A\tau^A$: different interventions which result in the same change in  $\alpha^{\eff}$ will lead to essentially the same impact on the disease. However, such an equivalence depends on the unknown value of $r$: two policies may lead to the same $\alpha^{\eff}$ for one value of $r$, but not for other values of $r$. This last point about how the impact of policies on the disease depends on the unobserved parameter $r$ is a central message of this paper. 

To underline the importance of targeting $\alpha^{\eff}$,  note that the basic reproduction number $R_0=\frac{\alpha^{\eff}N}{\gamma}$  is in fixed proportion to $\alpha^{\eff}$. From Table \ref{tab:alphaeffitaly}, we see immediately that to reduce $R_0$ to below $1$ in Italy on the policy date $T_2$, which is the goal of public policy, one needs to reduce $\alpha^{\eff}$  by an overall factor exceeding $4.40$. 

To this end, the following types of policy, labeled by $p$, can be implemented one at a time with varying efforts, or in combinations.  Table \ref{policies} provides the ad hoc benchmark parameters we use for expository purposes: Finding more realistic values is deserving of further study.

\begin{table}[width=.7\linewidth,cols=5,pos=h]
	\caption{Benchmark parameters for the four single policies with maximal effort. } 
	\label{policies}
	\centering
	\begin{tabular*}{\tblwidth}{@{} C|CCCC@{} }
		\toprule
		&$p=1$ &$p=2$ Social &$p=3$ Protective&$p=4$ \\&Isolation&Distancing& Garments&Hygiene\\
		\hline\hline 
		$v^O_p$&$f\phi=(0.9)(0.9)$ & 0.7 & 0.85 & 0.1 \\
		\hline 
		$v^A_p$& 0 & 0.7 & 0.85 & 0.1 \\
		\hline
		$e_p$ & 1 & 0.7 & 0.95 & 0.5\\
		\hline 
		$\alpha^O_p/\alpha^O$& 0.19 & 0.51& 0.19 & 0.95 \\
		\hline 
		$\alpha^A_p/\alpha^A$& 1 & 0.51 & 0.19 & 0.95 \\		
		\bottomrule
	\end{tabular*}
\end{table}

\begin{enumerate}
	\item {\bf Isolation of Infective Patients:\ }
	This type of policy ($p=1$) aims to prevent actively infectious cases from encountering susceptible people, and directly targets the close contact parameter $z^O$. Such a policy does not affect asymptomatic people, nor ordinary carriers that are not confirmed cases. We suppose that the maximal effect on any one confirmed carrier is to reduce their close contact rate by a fraction $f$. Thus the maximal effect vector of this policy is $v_1=(f\phi^O,0)$, and with effort $e_1$ its effect on the transmission parameters is 
	\begin{equation}
	\label{policy1}
	(\alpha^O,\alpha^A) \longrightarrow  (\alpha^O_1, \alpha^A_1) =\left((1-e_1f\phi^O)\alpha^O,\alpha^A\right)\  .
	\end{equation}

	\item{\bf{Social Distancing:\ }}
	Social distancing ($p=2$), such as a policy that requires keeping at least 2m distance in public spaces, can be targeted at identifiable sub-populations, or applied fairly across the general population. A strategy that targets the general population equally will have equal impact on the close contact fractions $z^O,z^A$ for both symptomatic viral carriers and asymptomatic viral carriers. If  implemented with effort $e_2$ and maximal effect vector $(v_2, v_2)$, the policy leads to 	
	\begin{equation}
	\label{policy3}
	(\alpha^O,\alpha^A) \longrightarrow (\alpha^O_2, \alpha^A_2) =\left((1-e_2v_2)\alpha^O,(1-e_2v_2)\alpha^A\right)\  .
	\end{equation}

	\item {\bf{Protective Garments:\ }}
	The wearing of personal protective equipment (PPE) ($p=3$), including gloves,	gowns, masks, face shields and eye protection, can be applied to the general population, and attempts to reduce the transmission probabilities $\tau^O,\tau^A$ equally. If  implemented with effort $e_3$ and a maximal effect vector $(v_3,v_3)$, this policy leads to
	\begin{equation}
	\label{policy2}
	(\alpha^O,\alpha^A) \longrightarrow (\alpha^O_3, \alpha^A_3) = \left((1-e_3v_3)\alpha^O,(1-e_3v_3)\alpha^A\right)\  .
	\end{equation}
	Some studies are helpful for determining $v$: For example, \cite{li2006vivo} claim that the efficiency of surgical masks is 95\%, compared with 97\% for N95 masks.

	\item{\bf{Hygiene:\ }}
	Infection via contaminated ``fomites''  (i.e. inanimate surfaces or objects), where active virus is absorbed from surfaces, has been considered an important mode of COVID transmission. Cleanliness ($p = 4$), particularly frequent handwashing and disinfecting surfaces, is the most important way of reducing spreading by viral contamination of fomites. 
	If implemented with effort $e$ and maximal effect vector $(v_4,v_4)$, a cleanliness policy has the following effect
	\begin{equation}
	\label{policy4}
	(\alpha^O,\alpha^A) \longrightarrow (\alpha^O_4, \alpha^A_4) = \left((1-e_4v_4)\alpha^O,(1-e_4v_4)\alpha^A\right)\  .
	\end{equation} 
	Recent studies summarized in \cite{Mondellietal_fomites_2020} have cast doubt on the overall importance of fomite transmission compared to aerosol transmission, and suggest that the maximal vector for this policy is small.
	
	
\end{enumerate}	
	
\begin{remarks}
	An effective vaccine is potentially the most powerful intervention tool: It acts directly on the immune system of susceptible individuals to reduce the infection probability $\tau$ to near zero. The SEAOR framework is inadequate to address vaccination: More preferable is to adopt a SVEAOR variation of our model that we do not pursue here. 
\end{remarks}

\subsection{Single Strategy}\label{sec:policy}

In this section, we use the calibrated base model for Italy to analyze the effect of implementing a single strategy. We  consider two distinct policy strategies, implemented singly: (a) isolation of all confirmed symptomatic patients (strategy $p=1$),  (b) protective garments for the general population (strategy $p=3$). We will demonstrate the important point that the effectiveness of a policy that applies equally to the entire population, as in case (b), does not depend on $r$. In contrast, the effectiveness of a policy that targets only confirmed active cases, such as isolation, can not be predicted without knowing $r$, and as we will see, other parameters. We therefore focus on the effect of these policies for different values of the asymptomatic rate $r$. 

\subsubsection{Effect of Isolation}\label{sec:eL}

Optimally, let us suppose that when fully implemented, each confirmed case reduces their overall transmission rate by $f=90\%$ in Eq \eqref{policy1}.

\begin{figure}[h!]
	\begin{subfigure}{.5\textwidth}
		\centering
		\includegraphics[width=0.95\linewidth]{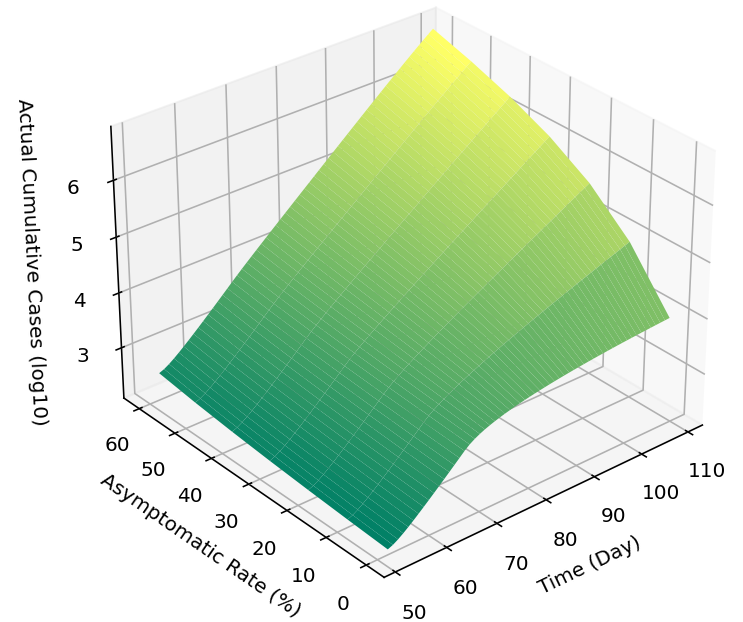}
		\caption{Actual Cumulative Cases}
	\end{subfigure}%
	\begin{subfigure}{.5\textwidth}
		\centering
		\includegraphics[width=0.95\linewidth]{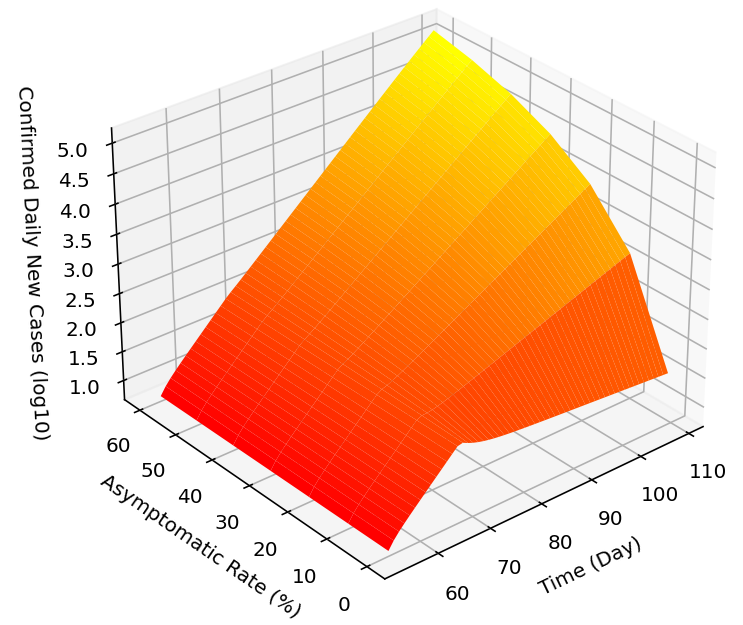}
		\caption{Confirmed Daily New Cases}
	\end{subfigure}	
	\caption{ITALY: The model prediction showing the effect of the maximal isolation policy if implemented on the policy time $T_2$ (March 9, 2020).  Here we fix the pre-policy ratio  $\alpha^A = 4\alpha^O$, the transmission reduction factor $f=0.9$ and the confirmation factor $\phi^O=0.9$, and plot over the period $[T_1,T_3]$ for varying $r$.  The first graph shows the actual cumulative cases consisting of all symptomatic and asymptomatic infections, and removed patients. The second graph shows confirmed daily new cases.}
	\label{fig:isor}
\end{figure}

The overall effectiveness of isolation on the disease itself is determined by the change in $\alpha^{\eff}=r\alpha^A+(1-r)\alpha^O$. This however depends on several other undetermined parameters, specifically $r$ and the ratio $\rho=\frac{\alpha^A}{\alpha^O}$:
\begin{equation}
\label{policy6}
\alpha^{\eff}=r\alpha^A+(1-r)\alpha^O \longrightarrow (1-r+r\rho)^{-1}\left[(1-r)(1-f\phi^O)+r\rho\right] \alpha^{\eff}. 
\end{equation}
Figure \ref{fig:isor} shows the $\log_{10}$ of the actual cumulative cases and confirmed daily new cases predicted by the model for Italy during the entire pre-post period $[T_1,T_3]$, under the maximal isolation policy. These graphs assume that before any policy $z^A=4z^O, \tau^O = \tau^A $, and hence  $\rho=4$. With these parameters, what appears to be a strong policy measure will fail outright if the asymptomatic rate $r$ exceeds about 10\% and of course the results will be even worse if the effort parameter is $e<1$ . 

Much more can be deduced from the right hand graph shown in Figure \ref{fig:isor}. As a consequence of Proposition 2, the isolation policy implemented with {\em any other combination of parameters} $e',f',\phi',r',\rho'$ that satisfy
\begin{equation}
(1-r'+r'\rho')^{-1}\left[(1-r')(1-e'f'\phi')+r'\rho'\right]= (1-r+2r)^{-1}\left[(1-r)(1-0.81)+2r\right]
\end{equation}
will have the identical time-development as the curve with rate $r$. 

\subsubsection{Effect of Protective Garments}\label{sec:eF}
\begin{figure}[h!]
	\begin{subfigure}{.5\textwidth}
		\centering
		\includegraphics[width=0.95\linewidth]{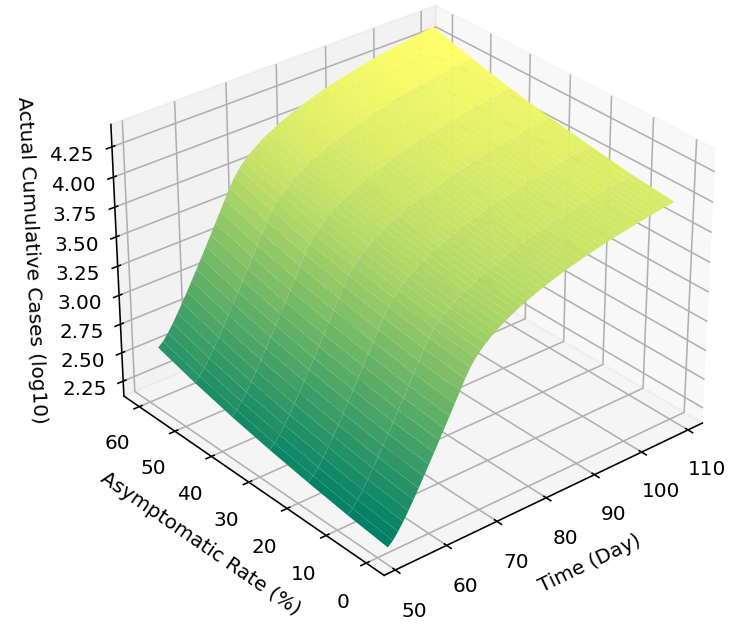}
		\caption{Actual Cumulative Cases}
	\end{subfigure}%
	\begin{subfigure}{.5\textwidth}
		\centering
		\includegraphics[width=0.95\linewidth]{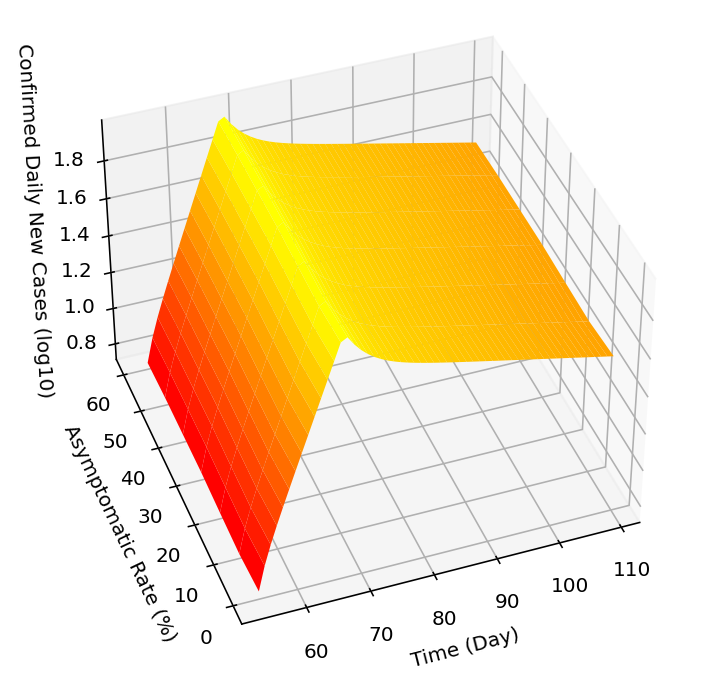}
		\caption{Confirmed Daily New Cases}
	\end{subfigure}	
	\caption{ITALY: The model prediction showing the effect of the maximal protective garments if implemented on the policy time $T_2$ (March 9, 2020).  Here we fix the policy factors $v_3=0.85, e_3 = 0.95$, and show that the pandemic curve over the period $[T_1,T_3]$ is independent of $r$.  The first graph shows the actual cumulative cases consisting of all symptomatic and asymptomatic infections, and removed patients. The second graph shows confirmed daily new cases.}
	\label{fig:genr}
\end{figure}

Here we consider the effect of a nation-wide policy of mask wearing where for definiteness we suppose there is a maximal effect $v_3 := v^O_3 = v^A_3 =0.85$, and a degree of effort $e_3 = 0.95$. In this setting, the result does not depend on $r$ or $\rho$. 
Figure \ref{fig:genr} shows the $\log_{10}$ of the actual cumulative cases and confirmed daily new cases predicted by the model for Italy during the entire pre-post period $[T_1,T_3]$, under the mask wearing policy. Since $\alpha^\eff\rightarrow(1-e_3v_3)\alpha^\eff$, we see that under all variations of the model with $v_3=0.85, e_3 = 0.95$ the pandemic is brought under control, with $R_0\sim 0.1925*4.40=0.847$.

\subsection{Effect of Other Single Policies}\label{sec:eg}
When the general population keep social distance (strategy $p=2$), the effect is similar to protective garments in that $\alpha^A,\alpha^O$ are changed by equal fractions $(v_2, v_2)$, even though it targets $z^A,z^O$ instead of $\tau^A,\tau^O$. This similarity also holds for policies of improved hygiene, if applied equally across the general population. If we combine these three strategies into a single policy we call {\em General Personal Protection}, then the logic of the previous analysis for protective garments remains unchanged.

Very unlike general personal protection policies, the effectiveness of the policy of isolation of confirmed active cases depends very strongly on the difficult-to-observe parameters $r, \phi$ and $\rho$, and in particular will perform especially poorly if $r$ is large. Indeed, under conservative assumptions   $r=10\%$, $f\phi=0.81$, $\rho=4$, the maximal isolation policy fails completely to halt the pandemic. 

Contact tracing can be regarded as a policy that improves the effectiveness of isolation by identifying a larger fraction of infectious cases. In other words, it seeks both to increase the parameter $\phi^O>\phi$, and to identify a fraction $\phi^A>0$ of asymptomatic cases. All these additional cases would then be included in the implementation of isolation policy. Effectively then, isolation combined with improved contact tracing has a maximal vector $(v^O_1,v^A_1)=(f\phi^O,f\phi^A)$.

\begin{figure}[h!]
	\begin{subfigure}{.5\textwidth}
		\centering
		\includegraphics[width=0.95\textwidth]{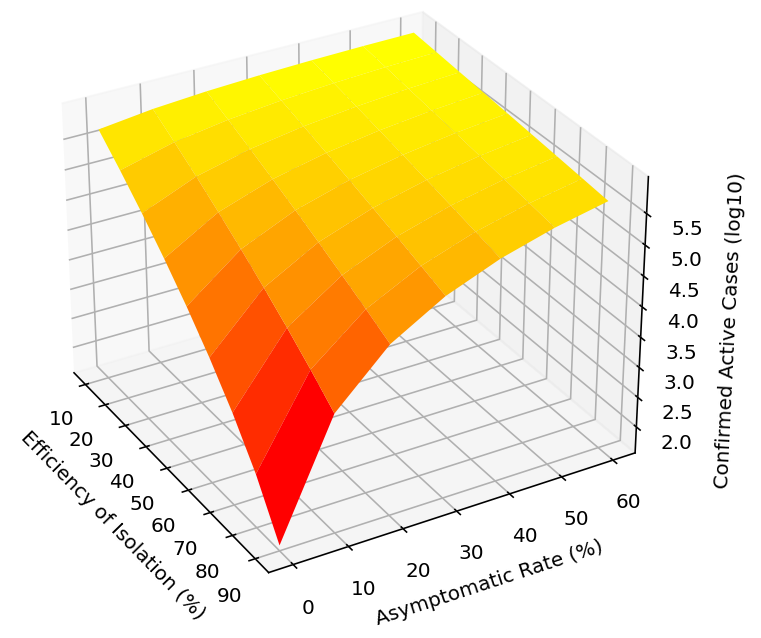}
		\caption{Isolation}
	\end{subfigure}%
	\begin{subfigure}{.5\textwidth}
		\centering
		\includegraphics[width=0.95\textwidth]{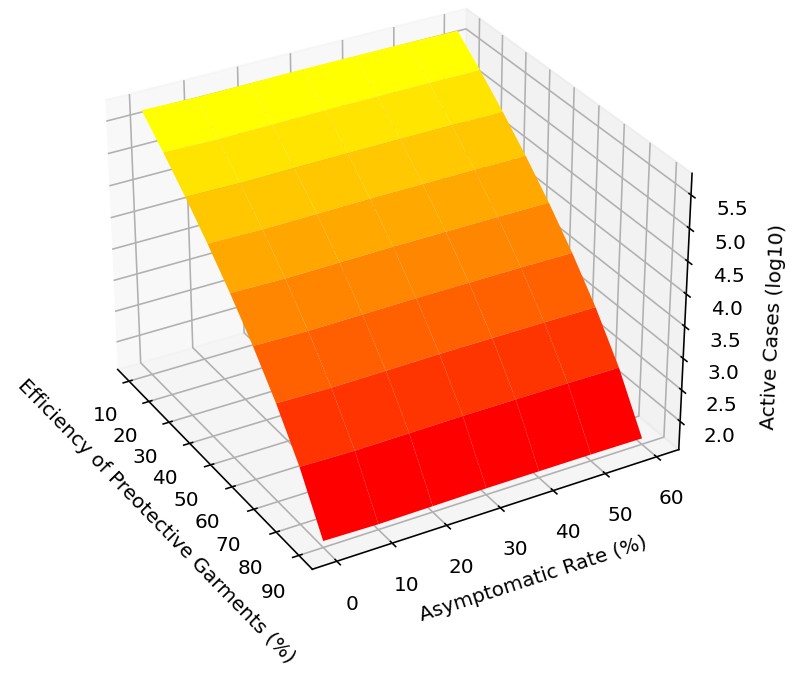}
		\caption{Protective Garments}
	\end{subfigure}	
	\caption{The effect of isolation and protective garments compared: These graphs show the logarithm of the active confirm cases in Italy on time $T_3$ corresponding to April 20, 2020. The left graph shows the dependence on the effort expended on isolation, $v_1e_1\in[0,1]$, and on the asymptomatic rate $r\in[0,0.6]$. The right graph  shows the dependence on the effort expended on protective garments, $v_3e_3\in[0,1]$, and on the asymptomatic rate $r\in[0,0.6]$ when $\phi = 1$ and $\rho = 4$. }
	\label{fig:genreff}
\end{figure}

Figure \ref{fig:genreff} shows how isolation and social distancing policies lead to very different outcomes for the pandemic. Under some reasonable assumptions on $r, \phi$ and $\rho$ the maximal isolation policy has very little impact on the pandemic, while the maximal protective garments policy eliminates the disease very quickly. 
 
\subsection{Combining Public Health Policies}\label{sec:compolicy}

Let us consider how the set of COVID mitigation strategies $p\in {\bf P}$ that are available to the policy maker can be implemented in combination, following the ``swiss cheese'' metaphor. First note that some care is needed to account for possible interference between the effects of different strategies. Under the following assumption, such interference effects have been eliminated:
\begin{ass}\label{dfn:alphae}[Independent policy assumption]
	The effect of the set of strategies $p\in {\bf P}$ when implemented in combination with efforts $e=(e_p)_{p\in {\bf P}}\in [0,1]^{\bf P}$ is to map the transmission parameters to new values:
	\begin{equation}
	\label{IPA}
	(\alpha^O,\alpha^A) \longrightarrow \left(\alpha^O \prod_{p\in {\bf P}}(1-e_pv^O_p),\alpha^A\prod_{p\in {\bf P}}(1-e_pv^A_p)\right)\ . 
	\end{equation}
	
\end{ass}

Let us take the point of view of the Italian National Health Authority that recognized the potentially devastating impact of the pandemic,  and implemented a remediation strategy effective on policy date $T_2$, March 9, 2020. The best  data available at that time indicates that the pandemic has a daily exponential growth rate $\lambda_+\sim 0.1999$, and an effective reproduction rate $R_0\sim 4.40$. To bring the pandemic under control will therefore require extreme measures: we suppose the authority aims to reduce $R_0$ to a value less than $0.8$, ensuring a reasonably quick resolution of the breakout.  Supposing that the authority fails to recognize the presence of asymptomatic carriers, we now study how their strategy  fails to produce a desirable outcome. 

Based on these assumptions, their first line of attack will be to expend maximal effort $e_1=1$ to identify and isolate all known active covid cases, following standard epidemic management policy. Because isolation ($p=1$) alone is insufficient to bring $R_0$ below 0.8, the authority needs to also consider more general strategies. To  this end they have identified three additional policies ($p=2,3,4$): social distancing (including shutting down some businesses and enforcing distancing rules), mask wearing and improved hygiene (including widespread use of hand sanitizer). We note again that if applied to the general population, these three policies have a similar impact on $R_0$ and can be combined into one policy which we call general personal protection (GPP). 

Assuming the benchmark parameter values shown in Table \ref{policies}, if the three policies are applied to the general population and implemented with efforts $(e_2,e_3,e_4)$, then following \eqref{IPA}, their collective maximal and partial effects will be multiplicative factors 
\begin{equation}
(1-v_{\GPP}):=\Pi_{p \in\{2,3,4\}} (1-v_p),\quad (1-e_{\GPP}v_{\GPP}):=\Pi_{p \in\{2,3,4\}} (1-e_pv_p ),
\end{equation} 
on both $\alpha^O,\alpha^A$. 

\begin{figure}[h!]
	\begin{subfigure}{.5\textwidth}
		\centering
		\includegraphics[width=0.95\linewidth]{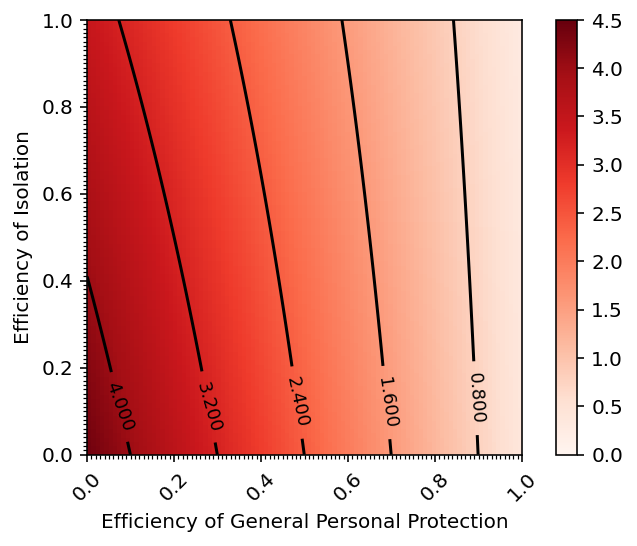}
		\caption{$R_0$ with two policies if $r = 40\%$.}
	\end{subfigure}%
	\begin{subfigure}{.5\textwidth}
		\centering
		\includegraphics[width=0.95\linewidth]{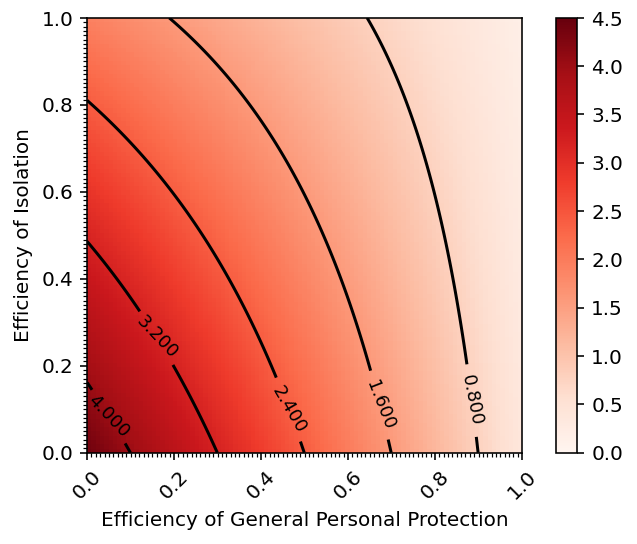}
		\caption{$R_0$ with two policies if $r = 10\%$.}
	\end{subfigure}%
	\caption{This contour plot shows the effective $R_0$ after policy date $T_2$ in Italy as a function of $e_1,e_{\GPP}$, when isolation is applied with effort $e_1$ and $v_1 = 0.8$, and general personal protection is applied with effort $e_{\GPP}$ and $ v_{\GPP} = 0.91$. The darkness of the red color denotes the value of $R_0$. Plot (a) shows results when $r=40\%$, (b) assumes $r = 10\%$.}
	\label{fig:comparecombine}
\end{figure}

Figure \ref{fig:comparecombine} shows contour plots of the achieved value of  $R_0$ under a combination of isolation with GPP for a variable level of $e_1,e_{\GPP}$ when $v_1 = 0.81, v_{\GPP} = 0.91$. The first assumes $r=40\%$ and the second assumes $10\%$. We see clearly from this that if $r=40\%$, achieving the desired value $R_0=0.8$ requires a far greater effort than the hypothetical proposed strategy. When $r = 40\%$, only a combination of GPP with $e_{\GPP}=90\%$ effort and maximal isolation policy can control the pandemic. A strict isolation policy combined with GPP with $e_{\GPP}=70\%$ effort that is sufficient to achieve $R_0=0.8$ if the authority mistakenly assumes $r = 10\%$, only achieves  $R_0=1.4$ if it turns out that $r = 40\%$.

\section{Discussion and Conclusions}\label{sec:diss}

The main aim of this paper has been to demonstrate why neglecting the existence of asymptomatic carriers is such a dangerous error for a disease like COVID-19. When planning health policy to control a pandemic such as this, it is easy to underestimate this effect, especially during the critical early stages. In hindsight, knowing now that $r$ is conservatively estimated to be larger than $30\%$, we can argue that indeed apart from parts of East Asia, most health policy in the world has failed outright largely because of this oversight.  

We have considered two distinct types of intervention policies: Isolation that can only target identified active confirmed cases and policies that target the general population.  Our main contribution is to demonstrate an essential difference between the two, namely that the former type of policy depends critically on a number of additional parameters that the latter do not depend on. The same value of $R_0=\alpha^\eff N/\gamma$ will arise from different specifications of the base model that cannot be distinguished with daily new case data, but these observationally equivalent base model specifications respond very differently to a policy change that involves isolation of confirmed cases. In model specifications where $r$ and $\rho$ are large, there will be hard to identify asymptomatic cases, and these cases will be on average more infectious than ordinary cases. In such circumstances, the effective $R_0$ achievable by a policy of isolation will be significantly underestimated.

A second conclusion is that many types of intervention that target the general population can be combined, with results that do not depend on $r$ and $\rho$ and are therefore very predictable. To the extent that $e_{\GPP}$ is sufficiently large, all combinations of policies involving general personal protection will be similarly effective in controlling the disease. 

With Proposition 2, this paper also makes a mathematical contribution by showing that under a natural assumption $\gamma^A=\gamma^O$, the reduced SEIR ODE model will provide almost complete information of the behaviour of the SEAOR model. This fact provides a very useful simplification for studying policy implications. 

Having a good estimate of the critical parameter $r$ in timely fashion is certainly essential to controlling COVID. Extensive testing accompanied by contact tracing, providing a large representative sample of the general population, is needed to adequately determine the impact of asymptomatic carriers. Testing and contact tracing also have the effect of raising the fractions $\phi^O,\phi^A$,  thereby improving the effectiveness and predictability of isolating known active cases. With extensive testing also comes more granular data that can identify the subpopulations that are most responsible for propagating the disease. When such data is used, more specific targeted interventions become practical and cost effective. Our methods can easily be generalized to analyze such targeted interventions. It is also important to reiterate that while $r$ may depend on the strain of COVID, its value does not depend on policy or social behaviour and is therefore stable over time and in different parts of the world. Thus scientists in every country need to follow carefully the world literature for information on this parameter.   

Many common sense aspects of public policy can be subjected to scrutiny using the techniques discussed in our paper. For example, we now know that mask wearing (strategy 3), would have been highly effective had it been widely implemented early in the pandemic. On the other hand, frequent hand washing (strategy 4) that was strongly advocated and adopted by the general population provided much less protection than was expected.  Restricting access to essential facilities such as grocery stores can backfire because social distancing requires as much space as possible. Finally, the shutting down of parks and open spaces in the early months of COVID was a weak strategy for controlling a disease whose transmission is dominated by aerosol and droplets.

\bibliographystyle{plainnat}

\bibliography{references_thesis}

\end{document}